# An Artificial Neural Network Algorithm to Retrieve Chlorophyll a for Northwest European Shelf Seas from Top of Atmosphere Ocean Colour Reflectance

Madjid Hadjal [1], Encarni Medina-López [2], Jinchang Ren [3], Alejandro Gallego [4], David McKee [1,5,*]

[1] Physics Department, University of Strathclyde, Glasgow, UK; madjid.hadjal@strath.ac.uk
[2] Institute for Infrastructure and Environment, School of Engineering, The University of Edinburgh, The King's Buildings, Edinburgh, UK; emedina@ed.ac.uk
[3] Department of Computing Sciences, Robert Gordon University of Aberdeen, UK; j.ren@rgu.ac.uk
[4] Marine Laboratory Aberdeen, Marine Scotland Science, Aberdeen, UK; alejandro.gallego@gov.scot
[5] Department of Arctic and Marine Biology, UiT the Arctic University of Norway, Tromsø, Norway; david.mckee@strath.ac.uk
* Correspondence: david.mckee@strath.ac.uk

**Abstract:** Chlorophyll-a (Chl) retrieval from ocean colour remote sensing is problematic for relatively turbid coastal waters due to the impact of non-algal materials on atmospheric correction and standard Chl algorithm performance. Artificial neural networks (NNs) provide an alternative approach for retrieval of Chl from space and results in northwest European shelf seas over the 2002-2020 period are shown. The NNs operate on 15 MODIS-Aqua visible and infrared bands and are tested using bottom of atmosphere (BOA), top of atmosphere (TOA) and Rayleigh corrected TOA reflectances (RC). In each case, a NN architecture consisting of 3 layers of 15 neurons improved performances and data availability compared to current state-of-the-art algorithms used in the region. The NN operating on TOA reflectance outperformed BOA and RC versions. By operating on TOA reflectance data, the NN approach overcomes the common but difficult problem of atmospheric correction in coastal waters. Moreover, the NN provides data for regions which other algorithms often mask out for turbid water or low zenith angle flags. A distinguishing feature of the NN approach is generation of associated product uncertainties based on multiple resampling of the training data set to produce a distribution of values for each pixel, and an example is shown for a coastal time series in the North Sea. The final output of the NN approach consists of a best-estimate image based on medians for each pixel and a second image representing uncertainty based on standard deviation for each pixel, providing pixel-specific estimates of uncertainty in the final product.

**Keywords:** Artificial Neural Network; Ocean Colour Remote Sensing; Modis Aqua; Chlorophyll a; Top-of-atmosphere; North Sea; Coastal waters

## 1. Introduction

Retrieval of ocean surface chlorophyll a (Chl, mg.m-3) is one of the key targets for ocean colour remote sensing (OCRS). The concept emerged in 1970 when Clarke et al. [1] observed the relationship between the colour of the ocean from aircraft measurements and the Chl concentration of the water. It has since been refined and applied to satellite sensors developed for the measurement of ocean colour resulting in continuous global daily coverage since 1997 [2]. The process consists of two main steps: first, removal of the atmospheric contribution to the top-of-atmosphere (TOA) signal measured by the sensor to produce a bottom-of-atmosphere (BOA) water leaving signal called the remote sensing reflectance (Rrs, sr-1) [3,4], and then, applying an algorithm to convert the Rrs spectral signal into ocean surface Chl.

Starting with the Sea-viewing Wide Field-of-view Sensor (SeaWiFS, 1997-2010), modern ocean colour satellite sensors have been equipped with infrared bands to support atmospheric correction. The so-called standard atmospheric correction algorithm (AC) [5] is based on the black pixel assumption that water absorption in the near infrared (NIR) is sufficiently high and backscattering sufficiently low that no light emerges from within the water column. The consequence of this is that at these wavelengths, measured TOA radiances can be assumed to result from atmospheric scattering only, and this signal forms the basis for extrapolation into the visible and removal of the atmospheric signal from measured TOA readings. It is well known that the black pixel approximation performs poorly in turbid waters, where particle backscattering can become a significant contributor to the NIR signal recorded by satellite sensors [6]. A number of alternative AC algorithms have subsequently been proposed [5-8] that aim to improve retrieval of water leaving signals in turbid or glint impacted waters. The black pixel approach remains in operation as the default option for processing NASA ocean colour data due to the fact that 90% of the surface of the ocean is not coastal. Despite significant efforts and progress to date, atmospheric correction remains problematic over turbid waters and there is no single, generally accepted method that is known to provide good quality BOA reflectances in such conditions.

Natural waters are classified into two optical water types following Morel and Prieur [9], with the optical properties of Case 1 waters being determined by phytoplankton and associated materials, and the optical properties of Case 2 waters being further influenced by non-covarying non-algal particles and coloured dissolved organic material (CDOM). Several algorithms have been developed to convert the water leaving signal into Chl, such as the blue-green algorithms OCx (Ocean Colour, using x bands, [2,10]) that were designed for Case 1 waters using the ratio between the blue and green wavebands. Derivations of this type of algorithm have been extensively studied recently by O'Reilly and Werdell [11]. These band ratio algorithms typically perform poorly in optically complex waters by overestimating the Chl as a consequence of the impact of other materials affecting the Rrs signals [12]. Coastal specific algorithms have been developed to overcome the problem [13-15]. Other approaches have been developed either for developing water-type specific variants of the blue-green algorithms [16] or by applying other classification schemes [17,18]. Further algorithms have been proposed for oligotrophic waters, such as the Colour Index [19] to improve predictions in oligotrophic areas, which is now regularly used for ocean colour algorithms when the Chl concentration drops below 0.2 mg.m-3. It is worth noting that Hu et al. [20] has recently shown that machine learning, in this case a technique based on support vector regression, has potential to improve retrieval of Chl for open ocean, Case 1 waters.

The focus of this work is the optically complex shelf seas off the northwest coast of Europe including the North Sea, Irish Sea, English Channel and western parts of the Baltic (see map in Figure 1, which also extends into the oceanic waters of the Northeast Atlantic). These shelf seas are socially and economically important and are subject to control through multiple international legislative agreements including the European Marine Strategy Framework Directive and the Water Framework Directive. As such, nations with territorial waters in this region are bound to implement effective monitoring programs to determine environmental status. These monitoring programs have traditionally been focused on shipboard surveys and deployment of moorings but there is growing interest in the potential to use satellite observations to extend the spatio-temporal coverage of observations. The key challenge is to ensure that satellite-derived Chl products are sufficiently reliable in order to be used for reporting against the legislative requirements. Current state-of-the-art algorithms for European North West shelf seas merge the OC5 [13] and CI [19] algorithms with different look-up-tables (LUTs) for OC5 processed by ACRI-ST through the GlobColour project or by Plymouth Marine Laboratory (PML), and are available on the Copernicus Marine Environment Monitoring Service (CMEMS, https://marine.copernicus.eu/ (accessed 9 May 2022). The performance of these algorithms is briefly assessed in this paper and used as a benchmark to compare against the performance of a new NN model.

Artificial neural networks (NNs) have been proposed to simulate biological neurons [21,22] and adapted to train single neurons to learn using perceptrons [23]. They consist of an input signal transformed into an output using an activation function with weights associated to each connection. Connections between multiple neurons and the definition of backpropagation of the error have been added later [24,25]. As a result of increased computation power availability, modern neural networks, especially deep learning networks, can contain up to billions of parameters and handle complex problems such as natural language processing [26]. NNs have initially been used in ocean colour for water classification [27]. The idea of using NNs for inverse modelling the light signal for Chl estimation emerged in 1994 [28]. Buckton et al. [29] applied NNs on modelled data and discussed the possibility of including non-light information in network training. NNs have been applied for Chl retrieval for Case 1 waters using either above surface measurements [30] or simulated data [31] or a mixture [32], with Keiner and Brown showing that NNs outperformed state-of-the-art algorithms at that time. Over optically complex waters, Schiller and Doerffer [33] used NNs with simulated Rayleigh-corrected reflectances, while D'Alimonte and Zibordi [34] applied the technique to a real coastal data set. In both cases the NNs returned promising results and / or better performance than state-of-the-art algorithms. NNs have been applied as operational products for case 2 waters constituents' retrieval for the Medium Resolution Imaging Spectrometer (MERIS, [35]) and the Ocean and Land Colour Instrument (OLCI) [36] radiometer sensors. Hieronymi et al. [37] have proposed a network trained on modelled data using the method developed in [38] applied to real satellite images with neural networks developed for classified water types being the key feature. [39] trained NNs for lakes with Sentinel 2 and 3 satellite data using above surface measurements showing good performance, while [40] have shown that NNs can outperform current state-of-the-art algorithms for Chl predictions in Chinese lakes. NNs have also been used to retrieve other parameters, such as photosynthetically available radiation [41], other pigments [42], Inherent Optical Properties (IOPs) [43], [44] and the spectral diffuse attenuation Kd [45]. Recently, NNs have been applied to retrieve surface temperature and salinity using TOA visible bands from the high resolution satellite Sentinel-2 [46,47]. Top of atmosphere signals have seldom been directly used by the OCRS community with only a few publications describing techniques relying on it [48,49], largely resulting from the fact that <10% of the signal in the blue is coming from the ocean for case 1 waters. It is clear that NNs have significant potential to improve retrieval of Chl and other important water quality and light field parameters from ocean colour signals in optically complex coastal waters. A review of the use of deep learning methods (i.e. more than one hidden layer) developed for Earth observation can be found in Yuan et al. [50] with a dedicated section to ocean colour.

Established, reliable and comprehensive data are essential for reporting against national and international water quality standards. Limited performance of existing ocean colour Chl algorithms in optically complex coastal waters is a major inhibiting factor in take-up of the technology by national environmental monitoring agencies. Many of the algorithms for coastal Chl are restricted in scope either geographically or through optical water type classification or by restricting application through extensive use of flags to eliminate the most challenging conditions, many of which are regularly found in northwest European shelf seas. This problem has persisted for over twenty years and there is little scope to believe that further development of blue-green reflectance ratio algorithms will significantly advance the issue [11]. However, satellite data have the potential to provide a degree of spatial and temporal coverage of Chl concentrations that is highly challenging or most likely impossible with alternative present-day in-situ observational technology, particularly in open sea or offshore areas [51]. The recent advancement of machine learning techniques suggests that it is time to develop a new framework for exploiting their strengths in OCRS. Whilst NNs have been discussed in the ocean colour literature since as far back as 1997 [29], they remain a new and unfamiliar territory for many researchers operating in OCRS and whilst there is a small but growing body of literature in this area (see above), understanding of the mathematical techniques involved and how to properly implement them remains confined to a relatively small element of the ocean colour community.

In this paper we aim to provide a detailed guide on how to develop a simple NN to derive Chl from OCRS data and to demonstrate that this approach is capable of providing estimates of Chl that are of similar quality to that provided by *in situ* sampling efforts. We illustrate the steps taken to identify appropriate architectures that optimise performance in terms of accuracy of Chl retrieval and focus on how this translates into ability to produce realistic mapped distributions of data. Building on work by Medina-Lopez et al. [47], we explore the potential to estimate Chl by applying NNs to TOA data directly. This approach would obviate the need for determination of appropriate ACs and effectively allows NNs to handle atmospheric signal impacts by inclusion of more bands than have been used before in the literature. Finally, we compare performance of resulting NNs against current state-of-the-art Chl products in terms of both data accuracy and data availability for a large set of matchup data covering northwestern European shelf seas and coastal waters.

## 2. Materials and Methods

*2.1. Study area and in situ data*

The study area for this work is shown in Figure 1 and extends from 25°W to 13°E and 48°N to 65°N, including samples from the North Sea, Irish Sea, English Channel, and the western Baltic Sea. These are predominantly optically complex Case 2 shallow shelf seas with many areas presenting high sediment and or CDOM loads [52] that influence OCRS signals either persistently or seasonally / episodically. There are also stations, e.g. in the NE Atlantic sector or from the northern North Sea, Arctic and Norwegian area, that are deeper and further from land which would satisfy the Case 1 classification but are clearly under-represented with respect to our data set (Figure 1a).

A Chl matchup data set covering the years 2002 - 2020 has been assembled from different sources of *in situ* samples: CMEMS (https://marine.copernicus.eu, accessed 9 May 2022); International Council for the Exploration of the Sea (ICES, www.ices.dk, accessed 9 May 2022); and data from countries included in the area that were directly provided by different institutions. For Danish marine waters, chlorophyll data were derived from the ODA database (DCE, 2021) and provided by the Department of Ecoscience, Aarhus University (Denmark). Data from the Norwegian and Barents Seas were provided by the Plankton Research Laboratory at the Institute of Marine Research (Bergen, Norway). For the Scottish waters and Stonehaven station, data were provided by Marine Scotland Science, Data from the waters of England and Wales were provided by the Centre for Environment, Fisheries and Aquaculture Science (CEFAS, https://www.cefas.co.uk/) and the Plymouth Marine Laboratory (PML, https://www.westernchannelobservatory.org.uk/, accessed 9 May 2022). The data set contains a mixture of Chl measurements produced using different methods [53], including: High Performance Liquid Chromatography [54] fluorescence [55], and spectrophotometry [56]. Usually samples with volumes typically ~1L will have been collected and filtered onto 25 mm GF/F glass fibre filters and frozen. The Chl pigments would generally have been extracted with 90% acetone and one of the methods specified above applied to measure their concentration. Additionally, this data set includes data from *in situ* fluorometry. Such a diverse data set naturally suffers from a range of complicating factors including differences between in vivo and extracted Chl concentration estimates due to factors such as solar quenching and also due to practical constraints such as pigment extraction efficiency. Round robin exercises have previously demonstrated uncertainties in HPLC concentrations up to 40% [57] and this more diverse data set could easily demonstrate errors of 50% or more depending on the measurement conditions [53]. Note that the CMEMS data set includes a large volume of data from ferrybox systems operating along the Norwegian coast. Unfortunately, this data set had to be eliminated from our analysis due to unresolved data quality issues. Other than these data from the Norwegian coast (approx. 60-65°N), no data were removed from the three data sets identified above. Approximately one million *in situ* Chl samples were available initially, but this number includes duplicates between different datasets, samples at different depths and Norwegian ferrybox data that were removed prior to averaging. The focus of this work is an attempt to establish satellite Chl products that provide equivalent quality data to that currently used by organisations like CMEMS and ICES. As such, we note that this validation data set is subject to unquantified and potentially significant uncertainty and that this should be considered in our analysis of satellite algorithm performance.

*2.2. Satellite data*

*2.2.1. MODIS Aqua*

The Moderate Resolution Imaging Spectroradiometer (MODIS) instrument on board the Aqua spacecraft has produced images since early July 2002. For this study, all MODIS-Aqua images available between 48°N and 65°N, 25°W and 13°E during daylight from July 2002 to January 2020 were downloaded as L1A products from the National Aeronautics and Space Administration (NASA) ocean colour servers (https://oceancolor.gsfc.nasa.gov/cgi/browse.pl, accessed 9 May 2022), using the R2018 calibration. The Aqua satellite has an ascending node orbit crossing the equator at 13:30. The MODIS sensor, with a swath of 2330 km and a pixel resolution of ~1 km at nadir observes approximately 80% of the specified area each day. In order to maximise the information content for the NNs to operate on, the following bands were saved for this study: 412, 443, 469, 488, 531, 547, 555, 645, 667, 678, 745, 859, 869, 1240 and 2130nm. Their characteristics can be accessed from the NASA website (https://modis.gsfc.nasa.gov/about/specifications.php, 10 March 2022). Bands 17-19 at 905nm, 936nm and 940nm were not processed due to their high correlation to cloud cover. Band 6 (1640 nm) has malfunctioned since 2006 [58] and therefore was not used for this study which leaves 15 bands. The inclusion of Bands 5 and 7 (1240 and 2130nm) follows the study of [59] who used these SWIR bands to perform enhanced atmospheric correction in coastal waters with MODIS Aqua. L1A files were downloaded, processed using l2gen and converted into L3 mapped files with a plate carrée projection using SeaDAS 7.5.1 following implementation of the NASA standard atmospheric correction using only the 'ATM FAIL' flag from the l2gen, with the "fudge option" set to 3. Images available from the same day were not merged in order to enable access to the temporal information and to provide optimal matchup conditions. This permissive approach, whereby flags that are usually applied by other data producers are not applied in this study, is intended to produce as broad a data set as possible in order to provide a test bed for assessing the potential for NNs to accommodate the most challenging optically complex waters. This has the added benefit of maximising the number of potential matchups which is the main limiting factor in NN development.

The total radiance measured at TOA by satellite sensors can be described as the sum of contributions from multiple physical effects:

$$L_t(\lambda) = L_R(\lambda) + L_a(\lambda) + L_{aR}(\lambda) + L_g(\lambda) + L_{wc}(\lambda) + L_w(\lambda) \tag{1}$$

with Lt the total radiance at TOA measured by the sensor. The terms on the right hand side of Eq. 1 are TOA radiances due to: LR total Rayleigh scattering by air molecules, La scattering by aerosols only, LaR aerosol-Rayleigh scattering, Lwc whitecaps and foam, Lg sun glint, and Lw the water leaving radiance. Three different reflectances were obtained as ouputs of l2gen process. The TOA reflectance, Rhot:

$$\text{Rhot} = \frac{\pi \cdot L_t}{F_o \cdot \mu_o} \tag{2}$$

The Rayleigh corrected reflectance, Rhos:

$$\text{Rhos} = \frac{\pi \cdot \left( \left( \frac{L_t}{t_{gsen} * t_{gsol}} \right) - L_r \right)}{F_o \cdot \mu_o \cdot t_{sen} \cdot t_{sol}} \tag{3}$$

The BOA remote-sensing reflectance, Rrs:

$$R_{rs} = \frac{L_w}{E_d} \quad (4)$$

with F0 the extra-terrestrial irradiance, µ0 the cosine of the solar zenith angle, $t_{gsen}$ and $t_{gsol}$ the solar to sensor and surface to sensor gaseous transmittances, $t_{sen}$ and $t_{sol}$ the solar to sensor and surface to sensor diffuse transmittances and Ed the downwelling radiance at the sea surface. While Rhot or Rhos can be used for quasi true colour image generation, Rrs is the apparent optical property used for most of the ocean colour algorithms.

The fourth-order polynomial ocean colour algorithm designed for case 1 waters for MODIS Aqua sensor (OC3M, [2]) was applied to available Rrs from MODIS Aqua matchups following equation 5:

$$\log_{10}(\text{chlor}_a) = a_0 + \sum_{i=1}^{4} a_i \left( \log_{10} \left( \frac{R_{rs}(\lambda_{blue})}{R_{rs}(\lambda_{green})} \right) \right)^i \quad (5)$$

with $a_0$ = 0.2424; $a_1$ = -2.7423; $a_2$ = 1.8017; $a_3$ = 0.0015 and $a_4$ = -1.2280 and where Rrs blue is the maximum Rrs value between 443 and 488 nm, and Rrs green is the Rrs at 547nm.

*2.2.2. Copernicus products*

A European Space Agency satellite product merging MODIS Aqua, SeaWIFS, MERIS, the Visible Infrared Imaging Radiometer Suite (VIIRS) and OLCI sensors created by the OC-CCI group is available for download on the CMEMS website. Rrs spectra of each sensor are realigned to MERIS wavebands at 412, 443, 490, 510, 560 and 665 nm and provides a daily merged and bias-corrected product for European shelf seas waters. It is available from 1998 to the present period. The NASA standard and polymer atmospheric corrections (respectively [3,7]) are applied depending on the sensor and area of study. One surface Chl product is available for download on the CMEMS website based on this merged daily Rrs product, named OC5-CCI, and uses the OC5 algorithm [13] in coastal waters. This algorithm was developed by IFREMER in collaboration with PML. It is available as a daily observed product and applies extra masking in certain conditions to avoid failure of the algorithm in coastal waters (see the latest Quality Information Document for this product, [60]). This algorithm will be referred to here as OC5-PML.

Another product is available from CMEMS using a similar approach (OC5 algorithm), provided by ACRI-ST, known as the European Space Agency GlobColour project, and is available as a daily interpolated product, with a +/- 30 days sliding window to create "cloud free" surface Chl maps [61]. This version will be referred to as OC5-ACRI. Rather than using OC-CCI merged Rrs data, it averages chlorophyll from each sensor separately, and creates an averaged version from multiple sensors afterwards.

Both of these algorithms (OC5-PML and OC5-ACRI) use a similar method, applying the OC5 and OCI algorithms in optically complex and clear waters respectively, therefore relying on classification. The OC5 algorithm was initially designed with 5 wavebands at 412, 443, 490, 510 and 550 nm and was developed to work for complex Case 2 waters impacted by constituents other than Chl, such as Coloured Dissolved Organic Matter (CDOM) or Total Suspended Sediments (TSS, [13]). The 412 nm band is used to take into consideration the CDOM absorption and the 550 nm band alone as a normalised water leaving signal to highlight the impact of sediments in water. It was initially trained with English Channel data. In open ocean waters, OC5 returns values close to the blue-green OC3 or OC4 algorithms [2]. The Colour Index (CI) algorithm, developed by Hu et al. [19], is a three band algorithm (443, 555 and 670 nm) used for oligotrophic waters with very low values of Chl (<0.2 mg.m-3), where blue-green algorithms can lead to failure because of different effects such as the glint, stray light etc. It was trained with oligotrophic to mesotrophic matchups. CI is used if the value returned by OC5 is below a threshold, usually around 0.15 or 0.20 mg.m-3. If the output falls between 0.15 and 0.2 mg.m-3, an average between both OC5 and CI results is returned. OC5 products used by both algorithms were developed using their own specific LUTs and available wavebands. The OC5 algorithm cannot return values above 65 mg.m-3 due to its design being based on LUTs not including values greater than 65 mg m-3. OC5-PML (OCEANCOLOUR_ATL_CHL_L3_REP_OBSERVATIONS_009_067) and OC5-ACRI (OCEANCOLOUR_ATL_CHL_L4_REP_OBSERVATIONS_009_098) were downloaded from the CMEMS website in Aug 2020 and Feb 2021 respectively.

*2.3. Generation of matchups between in situ and remotely sensed observations*

The formation of matchup data sets between *in situ* Chl concentrations and satellite data must take into consideration both spatial and temporal components. For comparison convenience, the same conditions were used for all three products. The temporal window considered all *in situ* data sampled from the first 10 meters, between 8:00 and 16:00 GMT+0 as a candidate. If multiple *in situ* samples were available for a given pixel at different depths, the median value was saved as a unique pixel matchup, with a priority for duplicate removal given to data available from the larger ICES dataset to increase consistency. The OC5-PML and OC5-ACRI products were reprojected using the MODIS Aqua grid as a standard to allow comparison over the exact same pixels.

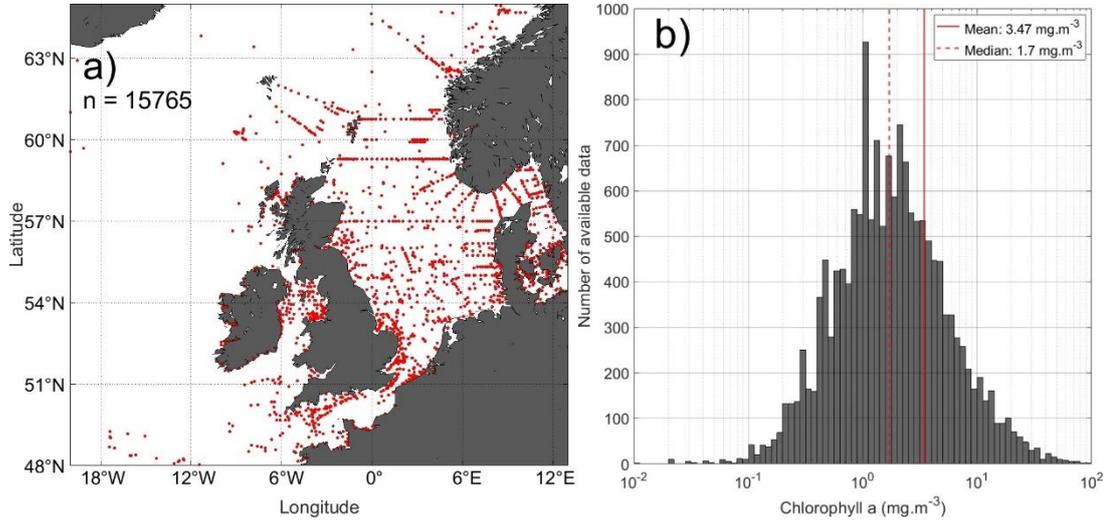

Figure 1: a) Map of all matchup points available for the MODIS-Aqua sensor, July 2002 to January 2020, 0-10m, averaged between 08:00 and 16:00. b) Histogram of the *in situ* samples.

The final matchup matrix consists of geo coordinates of the satellite pixel's centre, the three different reflectances (Rhot, Rhos, Rrs) at the 15 wavebands, and the median of the 0-10m Chl averaged between 08:00 and 16:00. Data duplicates from different sources (ICES, MSS and CMEMS) were removed before calculation of median Chl values. All the available matchups for the MODIS Aqua sensor are presented in Figure 1. The process has been repeated for the OC5-PML and OC5-ACRI algorithms, with different number of matchups available for each product (Figure 8). Final matchup numbers are determined by the details of processing for each algorithm tested later, but vary from 4757 for OC5-PML to 39331 for OC5-ACRI, with a total of 15765 matchups being available for the NN approach developed here using MODIS Aqua data, with 15 763 available when applying OC3 algorithm to MODIS Aqua Rrs.

The MODIS Aqua matchup data set is dominated by coastal waters, with the vast majority of the data sampled close to the coast. The final distribution is close to a normal distribution, with spikes at exactly 1 and 2 mg.m-3 and a median of 1.7 mg.m-3, which could come from sensor or human rounding. Over the 15765 matchups available for the MODIS Aqua sensor, 13246 are unique observations for different time and locations while 2519 matched at least two different MODIS Aqua images usually within a 1-hour interval due to the temporal window used and areas being seen twice by the satellite during the same day at these latitudes. These data were not merged and kept as unique matchups in order to add noise to the NNs as this has been shown to help NNs generalize [62]. There are approximately 1300 matchups per year between 2003 and 2006, 800 between 2007 and 2016 and less than 600 for each year afterwards. Seasonal coverage for the data set is not even, with approximately 1200-1400 matchups for months between February and September, less than 1000 for November and January, and less than 500 during December, mainly due to the increase in cloud cover during winter.

*2.4. Artificial neural networks*

*2.4.1. Neural network structure*

An artificial neuron consists of the application of an activation function associated with weight and bias that transforms an input signal coming from multiple sources into a predicted/estimated output. A feedforward neural network is a sum of neural network layers and composed of three main compartments. The first is an input layer including all the information available that could be useful to solve a problem, such as the 15 different wavebands available for this work. The second is a number of hidden layers each of which can have multiple neurons that are initialised with random weights and connected to adjacent layers. Finally, an output layer returns results based on the information produced by the final hidden layers. For each matchup, an error is calculated between the output produced by the network and the target, the *in situ* median of Chl. The error is back propagated and the weights of each neuron are adapted to minimise it until the network converges to a global minimum, if possible. It takes approximately 30 epochs to reach optimal performances.

For our study, all the layers are fully connected to the adjacent ones (Figure 2), with random weights used for initialisation. The input layer consists of the first 15 MODIS Aqua bands, using either Rhot, Rhos or Rrs. They were normalised using the min-max algorithm from Matlab's "mapminmax" function:

$$y = (ymax - ymin)\frac{(x - xmin)}{(xmax - xmin)} + ymin \qquad (6)$$

with the limits used being 0 and 1. The matchups dataset is randomly divided into three sets; 70% for the training set that will affect the weights evolution, 15% for the validation set that is used to stop training when the network is no longer improving; and the remaining 15% for the test set which are used to independently test the network. Backpropagation of the error measured between the *in situ* data and the prediction of the model is performed and training processes are repeated until the network converges and meets a global minimum, if possible, or a local minimum otherwise. The NN can fail to converge when the number of neurons or layers is inappropriate which is usually easy to diagnose as independent data (the test set) will show unrealistic behaviours such as over representation of specific concentration (predicting it as a line).

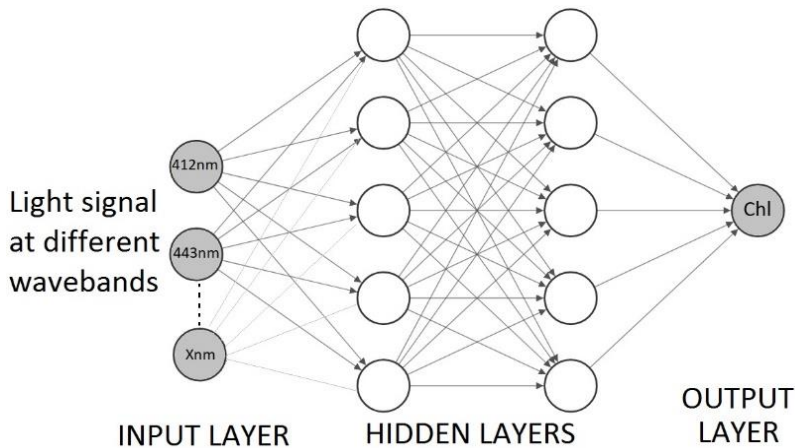

Figure 2: Simplified diagram of a fully connected multilayer perceptron neural network, showing for this example 2 hidden layers of 5 neurons each. Each arrow is associated to a weight, while the circles (neurons) are applying an activation function.

In NN development it is important to obtain an optimal architecture that produces good quality output data on both training and test sets and there is often a trade-off between the network complexity and the prediction accuracy. Selecting a small network structure may be computationally efficient but this can lead to under-fitting where there are too few connections to adequately resolve a complex signal. On the other hand, an overly big network is likely to be computationally inefficient and may introduce overfitting whereby the network uses a complex curve to predict a simple signal. In this case, the network may produce excellent results with the training set because it will remember the data set rather than learn from it, but it will give poor results with independent data. This is guarded against by testing the trained network against an independent test set. For this study, the whole process was conducted using Matlab R2020b's "fitnet" function from the deep learning toolbox to create the network and the "train" function to train the network, which is later applied to either the matchups or an image. We used the same number of neurons per layer each time (example in Figure 3), the Rectified Linear Units activation function for every node (y = x if x>0, else y = 0), and the Levenberg-Marquart function to minimise the error based on the Mean Squared Error. Scripts used in the production of data presented in this paper are available from the link in the reference section.

One of the more significant challenges of constructing a successful NN for this application is the need to be able to operate over a wide range of Chl concentrations. Our data set extends over ~3 orders of magnitude. In order to spread weights more evenly across the data set and following previous observations [63], the target (Chl) was log-transformed (Chl becomes log10(Chl)) prior to training and the inputs (15 reflectances) were normalised between 0 and 1 using the min-max method (Eq.5). Log transforming the target both improves network performance and prevents the network from returning negative values that would be unrealistic. Normalizing the inputs prevents the NN from relying too much on a dominant signal. Application of NNs to remote sensing images requires knowledge of the normalisation parameters used (the min and max values used for the normalization prior to training) and output values need to be back-transformed from log10 to obtain Chl concentrations.

2.4.2 Performance Metrics

Determining an optimal network architecture requires the selection of one or more performance metrics, which has previously been shown to be non-trivial for ocean colour applications [64]. In what follows we show results for two candidate metrics and consider their relative merits and demerits. The Pearson Correlation Coefficient metric R (Equation 7) is a common statistical descriptor for assessing algorithm performance but is known to be impacted by density fluctuations in the distribution of the dataset. [64] recommended use of the Mean Absolute Error (MAE - Equation 8) as being robust over several orders of magnitude, and as an absolute metric avoids being overly influenced by higher values. Here we describe this metric as Mean Absolute Difference (MAD) to reflect the fact that there are unknown errors in the *in situ* data set that mean it should not be considered as 'truth'.

$$R = \frac{\sum(M_i - \bar{M})(O_i - \bar{O})}{\sqrt{\sum(M_i - \bar{M})^2 \sum(O_i - \bar{O})^2}} \quad (7)$$

$$MAD = 10^{\wedge}\left(\frac{\sum_{i=1}^{n}|M_i - O_i|}{N}\right) \quad (8)$$

where M, O, and n represent the modelled value, the observation, and the sample size, respectively. Both M and O were converted into a log10 form prior to application. MAD was used to determine optimal network architecture, while R is reported to add statistical information to the matchup evaluations. An MAD of 1.8 as obtained with our study means that there is a relative measurement error of 80%. A smaller MAD value implies better performance of the algorithm.

**3. Results**

The data set we have assembled for northwest European waters is dominated by coastal waters, as illustrated in Figure 1a. Coastal waters present two important challenges for OCRS of Chl. The first is associated with degradation in the performance of standard blue-green reflectance ratio algorithms caused by absorption and scattering by non-algal particles and CDOM. This typically leads to overestimation of Chl by variable amounts which are both spatially and temporally dependent [16]. The second challenge is the impact of backscattering by non-algal particles causing non-zero water leaving radiance in the NIR, breaching the initial assumption of the black pixel atmospheric correction [65] and conducted to new strategies . This leads to production of incorrect Rrs values which in turn causes further breakdown of the blue-green reflectance ratio algorithms. Indeed, this failure of the standard AC in coastal waters would potentially affect the performance of a NN operating on Rrs values (see later). As a result, here we test NNs operating on both AC-corrected (Rrs), Rayleigh corrected (Rhos) and uncorrected top of atmosphere (Rhot) reflectances.

3.1. Identification of optimal network architectures

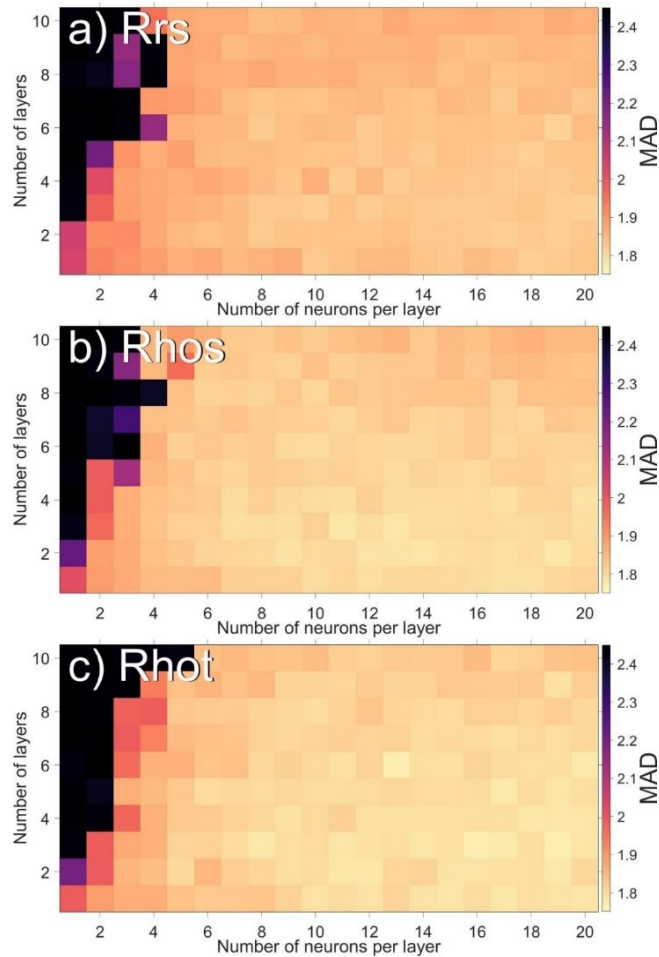

Figure 3: Mean absolute difference heat maps applied to the test set for neural networks operating on Rrs (a), Rhos (b) and Rhot (c).

Various strategies can be adopted for finding optimal network architectures. It is reasonably common practice to take at least the same number of neurons as inputs and to evaluate the impact of adding more hidden layers. Because we do not have an overly large dataset and since even reasonably priced modern computers have good performance characteristics, we opted to systematically explore the impact of selecting different network architectures. MAD scores were obtained for NNs operating on Rrs, Rhos and Rhot inputs. In each case, we tested NN architectures varying between 1 to 10 layers and 1 to 20 neurons per layer. We repeated the application of each architecture 10 times. The median MAD value of the 10 runs applied to the test sets (15% of the total data set) is displayed in Figure 3. Similar performances are obtained with the training sets (not shown). There is a general tendency to obtain higher MAD scores for architectures using between 6 to 20 neurons per layer, and less than 5 layers. However, identifying a truly optimal 'winner' for each input data type (Rrs, Rhos and Rhot) is probably not meaningful. Rather there are regions in this space where performance is broadly equivalent, and will be slightly different each time due to the effect of the random initialisation of the weights. In this case prediction results are similar once we have at least 6 neurons per layer, which could highlight that there may be elements of redundancy over the 15 bands available. For this study, we used networks composed of 3 layers of 15 neurons for the Rrs, Rhos and Rhot reflectances as they produce nearly optimal results without becoming overly computationally intensive. Choosing a higher architecture, say 8 layers of 20 neurons, was found to give better performance on the training sets (higher metrics) but poorer performance on the test sets. At least two layers were required to avoid underfitting issues which sometimes happened when a single hidden layer was used, but was not obvious from MAD metrics but was clear from visual inspection of plots. Differences in performance between test and training data sets can be a sign of overfitting, i.e. failure of the network to generalise. It is clear from Figure 3 that having too many hidden layers without enough neurons per layer generally degrades performance.

The heat maps in Figure 3 also reveal differences in the level of performance between NNs operating on different input sources. Interestingly, NNs operating on uncorrected TOA Rhot input data perform best, slightly better than Rhos, with Rrs showing the poorest performances. NNs operating on fully atmosphere corrected Rrs values produce higher MADs, though the differences do not appear to be very large and any of the three reflectance could still be used effectively. At first glance, it may seem surprising that uncorrected TOA reflectance inputs, Rhot, produce such apparently stronger results despite the atmospheric reflectance signal being present within the input. This will be examined in more detail later, but it should be realised from the outset that this effectively means that in this case the NN is having to account for the impact of atmospheric scattering and having to derive Chl for a wide range of coastal water types. On the other hand, Rrs and Rhos NNs are operating on reflectance data that is imperfectly corrected for atmospheric effects. As shall be highlighted later (Figure 5), both the full atmospheric correction and even the Rayleigh correction have potential to generate unphysical (negative) reflectance data, especially for turbid coastal waters. In these cases, the NN is effectively having to compensate for these AC errors and then derive Chl for optically complex waters. Taking this into consideration, it is perhaps less surprising that the TOA Rhot NN performs slightly better than the other two due to the loss on information during the AC process.

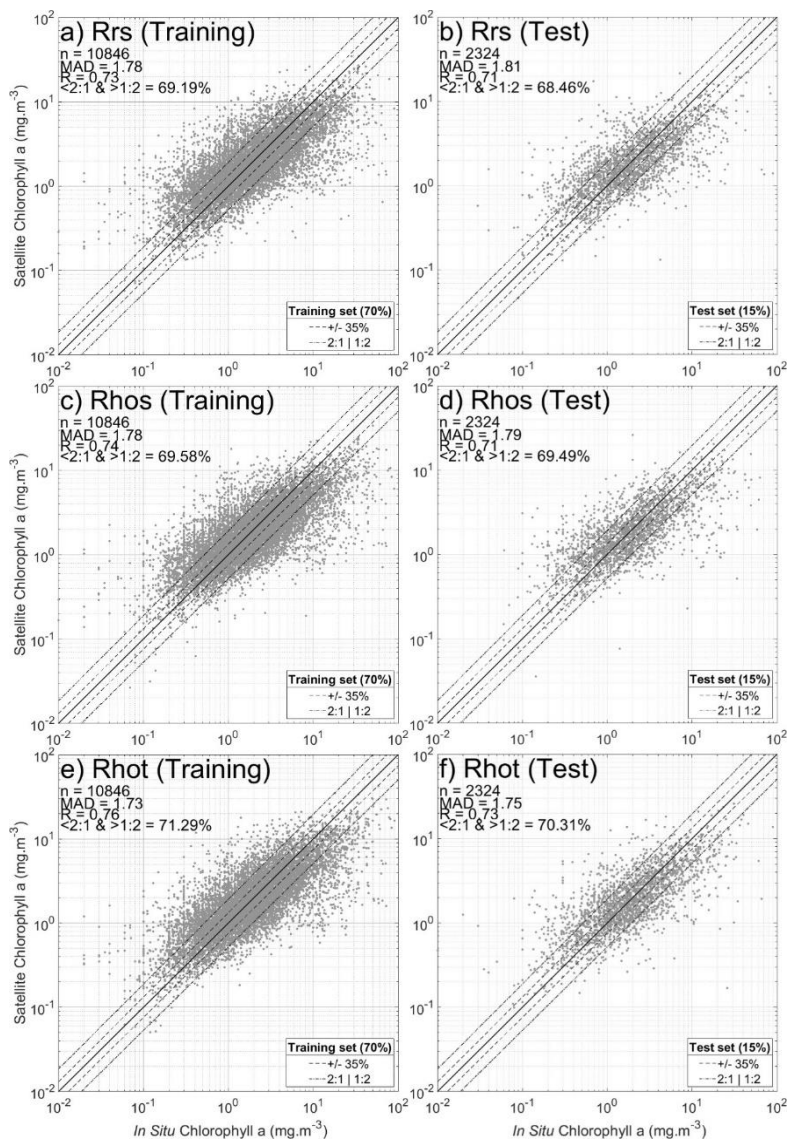

Figure 4: Neural network performances for an architecture based on 3 layers of 15 neurons each, using Rrs (a and b), Rhos (c and d), and Rhot (e and f) for the training set (70% of the total matchups) and the test set (15% of the total matchups) respectively.

Figure 4 shows the performances of the NNs for both the training (70%) and test set (15%) applied to the 15765 matchups available, the last 15% (1459 points) being used as the validation set (not shown). All three reflectances show similar performances. There is a slight tendency to overestimate low values (<1 mg.m-3) and to underestimate high values (>10 mg.m-3), possibly reflecting limits of representation in the training data set (not enough training data available for these ranges). Approximately 70% of the points fall between the 1:2 / 2:1 dashed line (a ratio of 2 between *in situ* samples and satellite estimation), close to the *in situ* error measurements for such a data set. The gap between training and test set performances is small, with Rrs showing higher differences than Rhot. Rhot achieves the best performance for the test set and at this stage is the candidate of choice for use in the rest of this publication. However, any of the reflectances could be used, as only some spectra were over corrected by the AC, and thanks to normalisation of data prior to training all show relatively good performance for a coastal data set.

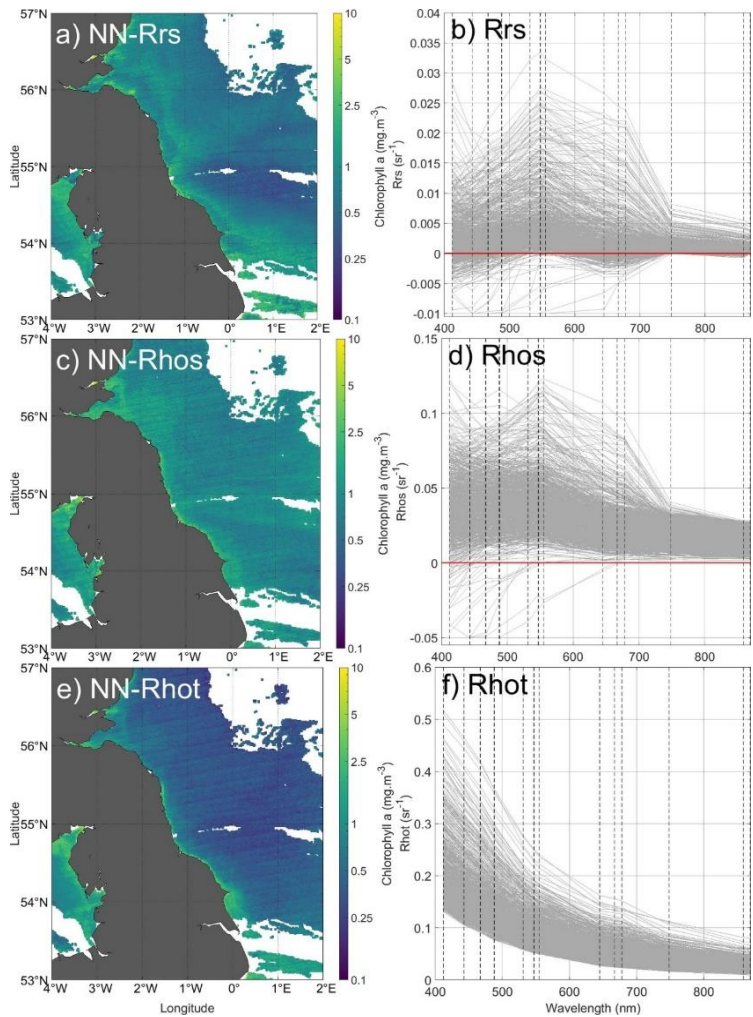

Figure 5: Winter image of the 31th of December, 2019, 13:05 from MODIS-Aqua, highlighting the atmospheric correction noise present near the Firth of Forth, western North Sea for the Rrs a) and Rhos c). Absent from the Rhot product e). 1000 spectra examples from the matchup data set, for Rrs b), Rhos d) and Rhot f). Only the first 13 of the 15 bands of MODIS Aqua available for this study are displayed with dashed lines.

To further support the choice of Rhot, Figures 5a, c and e show one example of a winter map that highlights the impact of atmospheric over-correction on NN performances for both Rrs and Rhos. It is visible as both dark patches some distance off the coast and as high values associated with image striping in the western North Sea. Features like these commonly occur in areas where turbidity is known to be high when using NN based on Rrs or Rhos, but are generally absent using Rhot. This can be illustrated by examining a subset of the available matchup spectra (Figure 5b, d and f). It is well known that the standard, black pixel atmospheric correction causes over-correction of Rrs spectra for turbid coastal waters, seen here most obviously as the occurrence of negative Rrs values (Figure 5b), but potentially being true even for non-negative data. Intriguingly, negative values are also found for a smaller number of Rhos data (Figure 5d), implying that even taking the preliminary step of applying the Rayleigh correction can sometimes be sufficient to produce unphysical data. None of these negative Rrs or Rhos values are realistic, and we suspect this is the main reason why the Rrs and Rhos NNs show slightly poorer performances, with the NN having to overcome this type of over-correction at the same time as determining the Chl signal. In contrast, Figure 5f shows TOA reflectance (Rhot) signals which are always physically plausible, even if they are still obviously impacted by the contribution of atmospheric scattering. MODIS Aqua striping effects seem to be more visible in winter compared to other seasons, which may highlight a link with high solar zenith angles and a need for removal prior to application of any algorithm [66]. Our results suggest that it is marginally easier for the NN to handle the uncorrected atmospheric scattering signal in Rhot than it is to undo imperfect atmospheric correction of Rhos and Rrs signals. The combination of higher statistical metric performances, the training and test set performances being closer, and observation of unrealistic oceanic coastal features possibly due to the failure of atmospheric correction all lead to the choice of Rhot as the preferred data source for NN development.

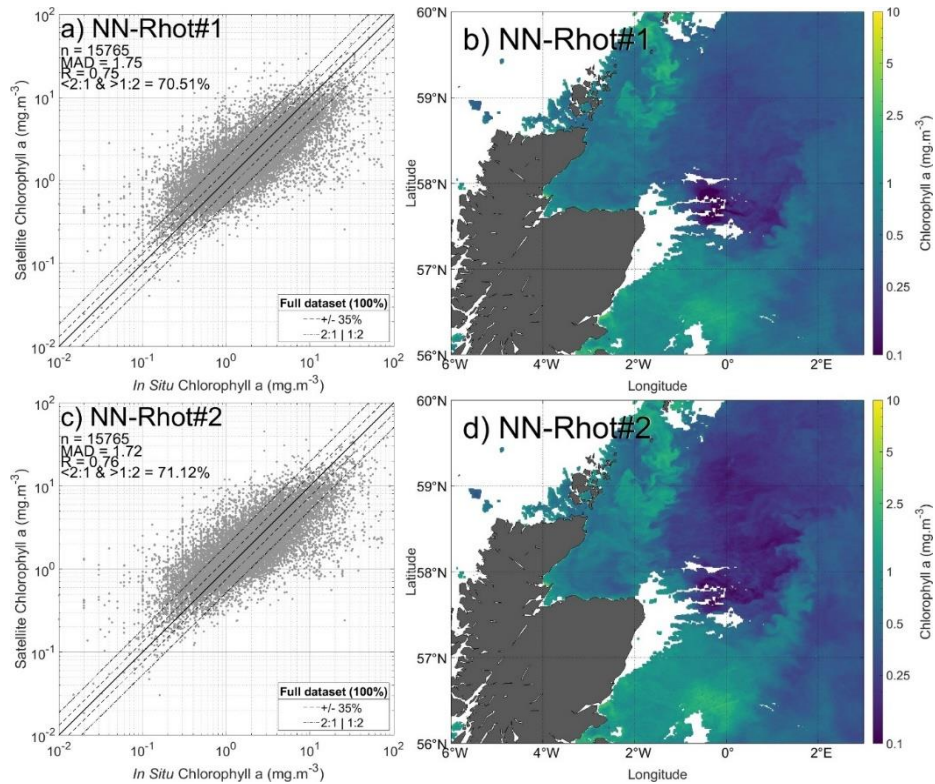

Figure 6: Two examples of the same architecture of a Rhot neural network using 3 layers of 15 neurons showing slight differences. a) and c) Neural network performances for the whole matchup data set. b) and d) Image of the 23rd of July 2019, 12:25 from MODIS-Aqua (same as image 7). Notice the difference around 0°E and 57.5°N coming from a coccolithophore bloom returning very low Chl values.

The nature of NNs is such that, each time a network is trained, it will produce a network that is specific to the training data set employed. Randomly re-sampling the available training data generates subsequent NNs with properties that are not exactly the same each time. Figures 6a and 6c display the results of two NNs using Rhot and that return similar yet slightly different performances. These differences are even more apparent from their respective images (Figure 6b and 6d) for the 23rd July, 2019 at 12:25 from MODIS Aqua, where a coccolithophore bloom near 57°N and 0°W is retrieved differently by the two NNs. This discrepancy is largely due to the fact that coccolithophore blooms are underrepresented in our matchup data set and performance of the resulting NNs is therefore heavily dependent on how many such stations are included in the associated training data sets. Elsewhere the two NNs produce images that are visually very similar. The obvious solution to this problem is clearly to attempt to expand the training data set through targeted sampling at sea. However, this is impractical in the short term. Instead, we must look to develop an approach that is more robust for any given training data set.

To this end and to minimise the impact of under-sampled features in our training data set such as coccolithophores blooms, we decided to use a standard NN architecture (3 layers of 15 neurons) but resample the training data set multiple times, generating multiple NNs that could subsequently be analysed to produce a single, hopefully convergent, median data product. This approach has the further merit of being able to provide a measure of product uncertainty through the standard deviations of the resulting distributions of Chl values for each pixel.

Conversely, the computational effort involved needs to be considered if the approach is to be used in an operational sense for image processing, so establishing an optimal number of NN iterations is essential. Figure 7 displays median values for the same image of the 23rd July 2019 using 10 (Figure 7a) and 100 (Figure 7b) networks. Visual inspection of these panels (and others – not shown - representing different numbers of iterations) suggest that an ensemble approach using the median of 10 NNs is sufficient to achieve convergence with a version merging 100 NNs. Figures 7c and 7d show corresponding relative standard deviations from the 10 and 100 iteration networks respectively, expressed as percentages relative to the median (rather than the mean). Again, there is broad consistency between these two images suggesting that 10 iterations is sufficient to capture the performance of the NN approach. The standard deviation of NN outputs varies across the image, reflecting variable confidence in NN output for each pixel. Figure 7e shows that this percentage uncertainty varies from ~7% to more than 100% with the vast majority of data falling between 10 and 30%. Looking at a worst case scenario, Figure 7f shows the impact of this uncertainty for a randomly selected single point inside the coccolithophore bloom near 57°N and 0.2°W where NN performance is worst, revealing up to an order of magnitude uncertainty.

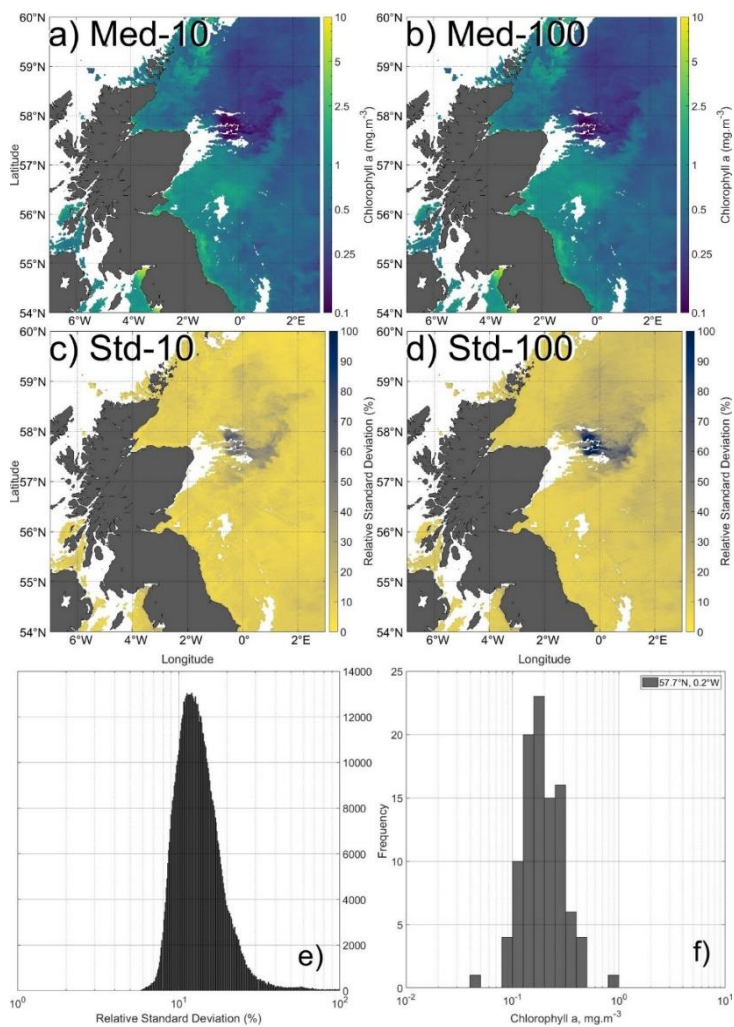

Figure 7: Chlorophyll prediction from the neural networks using a) 10 networks and b) 100 networks averaged (median). Relative standard deviation (relative to the median) for the same image expressed in percentage using c) the same 10 or d) 100 networks. e) Histogram of the relative standard deviation of the panel d) using 100 networks expressed in percentage. f) Histogram of a random point inside the coccolithophore bloom.

3.2. Algorithm performance evaluation

In order to establish a baseline for performance evaluation with current mainstream Chl algorithms, Figure 8 shows results for all the available matchups for a) OC3, b) OC5-PML, c) OC5-ACRI and d) MODIS Aqua data processed using the median of 10 NNs with the same 3 layers of 15 neurons architecture, using Rhot. The OC3 algorithm was applied to MODIS Aqua matchups Rrs obtained using the permissive approach and therefore shows a massive spread between measured and retrieved Chl, with a strong bias towards over-estimation. The OC3 algorithm was not designed for optically complex coastal waters, and initial NASA development of this algorithm relied on application of masks to eliminate pixels with obvious data quality issues; see https://oceancolor.gsfc.nasa.gov/atbd/ocl2flags/ to see when specific flags are applied at level 2 or 3 or the data. While some data follow the 1:1 line, most of the matchups are overestimated by the algorithm by varying amounts but up to by several orders of magnitude. This is likely due to the influence of turbid waters on both atmospheric correction performance and OC3 algorithm performance, which usually return very high Chl values over sediment plumes on images. This is not surprising as the algorithm was developed for clear open ocean waters and is not expected to perform well in turbid coastal waters, where a good proportion of the available matchups come from. Of course, it is also worth noting that this algorithm (and others with similar structure and performance in these waters) remains in common use by unwary end users who perhaps have less familiarity with the field, and who would potentially benefit from more robust guidance by data providers.

The OC5-PML algorithm shows a clear improvement compared to OC3, with overestimation restricted to a maximum of ~1 order of magnitude for Chl between 1 and 10 mg.m-3 and possibly a tendency towards underestimation at high concentrations. Importantly, only 4757 matchups were available for this product despite being based on the accumulation of data from between 2 and 5 satellites at any given time. This significant reduction in data availability comes from: 1) the OC5 algorithm itself which can only be applied in certain conditions (based on the signals at 412 and 550 nm), which removes a significant number of potential matchups; and 2) additional quality control flags which exclude more problematic waters such as coccolithophore blooms, very coastal or shallow waters, glint impacted areas, low sun angles, etc. It should also be noted that the maximum value allowed by the OC5 algorithm is 65 mg.m-3, which can be problematic for coastal waters where higher concentrations are possible.

The OC5-ACRI product (Figure 8c), despite using a broadly similar algorithm to OC5-PML, tells a different story, with many more matchups being available, but with significantly greater ranges of over and under-estimations. The increase in data availability is directly due to use of a +/- 30 days sliding window average rather than single direct observations. It is likely that the apparently stronger performance of the PML variant is achieved through use of additional flagging to remove poor quality data rather than actual improvement in algorithm performance per se. Conversely, while the OC5-ACRI product has a clear advantage of having almost 100% coverage for the area except in winter, increasing by ~3 times the number of available matchups, this appears to be achieved at the price of data quality. There is clear potential merit in using this type of merging approach at global scales that are dominated by case 1 waters where the algorithm may perform well, but these results suggest that there may be significant issues in coastal regions.

The proposed ensemble NN-Rhot product (Figure 8d), offers several advantages. The distribution is clearly better constrained towards the 1:1 line than either OC3 or the OC5-ACRI product (Figures 8a and 8c), and is somewhat tighter than the OC5-PML product (Figure 8b) with R of >0.75 vs 0.61 and MAD of <1.8 vs 2.1. To be noted, the slightly improved performances reaching MAD <1.73 come from the inclusion of both train and test set performances together, and similar performances to what have been shown earlier in Figure 4f should be expected, with the real MAD probably lying between 1.75 and 1.8. The NN approach produces more than twice the matchups available for OC5-PML (13246 "daily" matchups for the permissive MODIS Aqua approach against 4757), and it does not require application of further flags to eliminate optically complex waters or outliers to reach similar performances. Very importantly, the NN performance is achieved without requiring the application of any atmospheric correction, even though this contains a wide variety of optically complex water conditions. The NN product is far from perfect and there is evidence of a tendency to over-estimate at low concentrations and vice versa, with the range of error remaining at approximately one order of magnitude on a per pixel basis. However, two thirds of the NN data lie between the 1:2 and 2:1 lines, broadly reflecting the level of performance that can be attained for *in situ* measurement of Chl using the diverse methods used to generate the training data set. It is worth noting that both the PML and ACRI algorithms could potentially return higher performances if retrained using this specific matchup dataset. That said, the NN approach appears to offer a useful combination of high quality performance and maximal data availability.

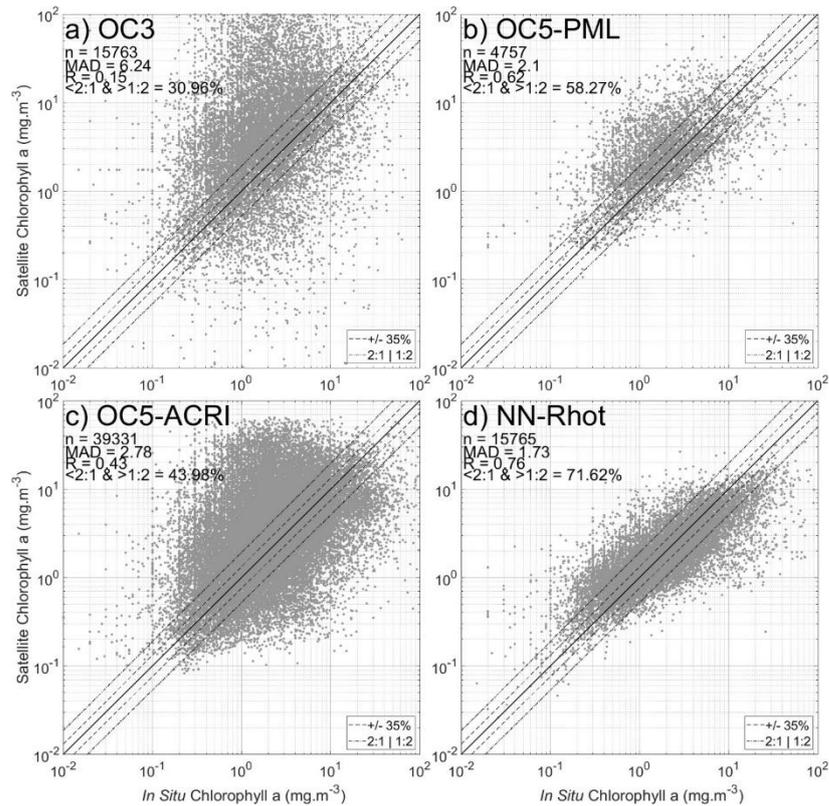

Figure 8: Performances over the matchup dataset for a) OC3, b) OC5-PML, c) OC5-ACRI and d) NN-Rhot - the median of 10 random neural networks using Rhot with MODIS Aqua data.

To avoid data availability bias between the different products used above as comparators, we have repeated the analysis but this time have restricted the comparison data set to matchups that are available for all four algorithms. As the most restrictive algorithm examined here, this second data set is largely constrained by the flagging procedures adopted by the OC5-PML algorithm. However the process of establishing clean matchups for each algorithm means that there are fewer common matchups than were originally available even for OC5-PML. Therefore, only samples commonly available for OC5-PML, OC5-ACRI and the permissive MODIS Aqua dataset are shown in Figure 9. The vast majority of OC5-ACRI matchups following this approach should come from direct observations due to the requirement of data being available from the MODIS Aqua and OC5-PML daily observations. 3896 points met the requirements of being available from all products at the same time. Performances of OC3 and OC5-ACRI products are improved (with a very similar distribution). While the OC5-PML algorithm returns similar metrics (MAD of 2.09), the NN-Rhot approach benefitted from flagging data following OC5-PML approach with an improved MAD below 1.7. For reference, the MAD obtained for case 1 waters algorithms like GSM [67] OCI or OC3 using the SeaBASS dataset (SeaWiFS Bio-optical Archive and Storage System, [68]) reach ~1.6 [64], with best and worst performances reached over oligotrophic and eutrophic waters respectively from the GSM algorithm (MAD of 1.47 and 2.05). It is notable that the NN approach developed in this study appears to achieve performance metrics for optically complex coastal waters that are comparable with standard algorithm performance in Case 1 waters.

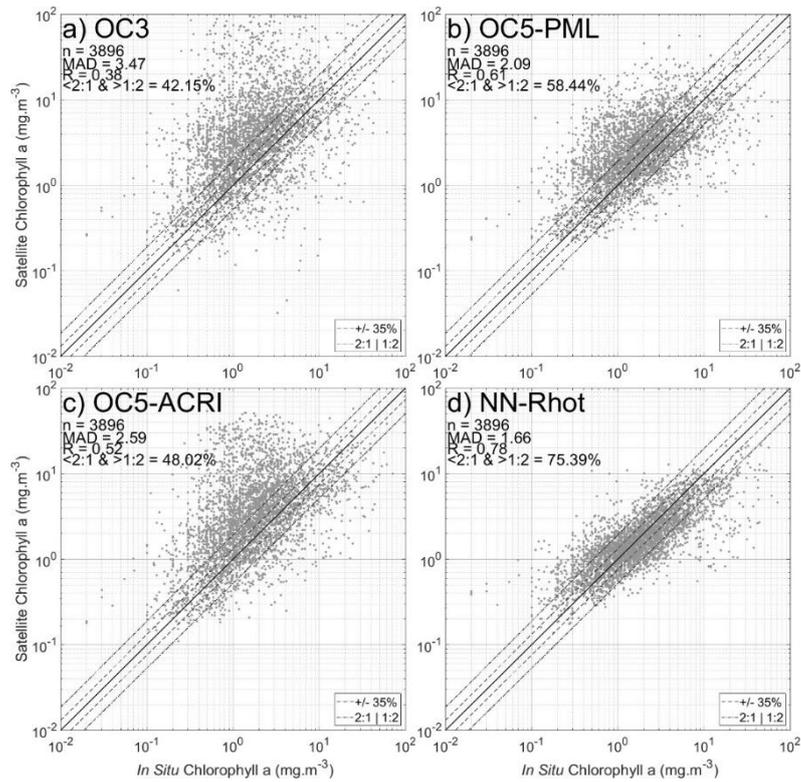

Figure 9: Performances over the exact same matchup dataset for a) OC3, b) OC5-PML, c) OC5-ACRI and d) NN-Rhot - the median of 10 random neural networks using Rhot with MODIS Aqua data.

Further comparative evaluation of algorithm performance is achieved through analysis of images from spring (Figure 10) and winter (Figure 11). Figure 10 shows an example of a spring day at the start of the spring bloom season (20th of April 2005). In general, OC3 produces higher maximum values, well beyond the top end of the colour scale used in the plots, with a maximum of ~4300 mg.m-3. In comparison, the OC5-ACRI, OC5-PML and NN-Rhot products reach maximum values around 62-65 mg.m-3 due to OC5 not allowing any value above 65 mg.m-3 while no threshold was defined for the NN. OC3 and OC5-ACRI display broadly similar results across the scene including a patch of high Chl values off the east coast of Scotland which could potentially be an artefact of image merging. This observation is consistent with the previous section and Figure 9 where both products returned similar distributions. OC5-PML and the NN are in broad agreement with relatively small differences between them in the North Sea area of this image, the main difference being in the Baltic Sea where very coastal waters are returned as high Chl values by the PML product with no particular feature seen from the NN. It has been previously observed that the PML product overestimates Chl in the Baltic part of the image due to presence of CDOM [69]. Another area where we can spot differences, this time between the NN and the blue-green algorithms, is in the NE Atlantic where the NN produces higher values than any of the other products which generally agree with each other by returning values below 0.25 mg.m-3, largely due to the fact that the OCx, GSM and CI algorithms are applied for this area and there is generally only limited difference in performance for these relatively clear waters. In this case, it is likely that the OCx products are performing well and the NN would benefit from inclusion of additional training data from oligotrophic waters. This difference over an under-sampled area like the NE Atlantic highlights a need for the matchup dataset to contain more open case 1 waters and re-emphasises that this is a data driven approach. NNs do not have the capacity to make realistic estimations for under-sampled scenarios (only ~3% of the current dataset comes from the NE Atlantic).

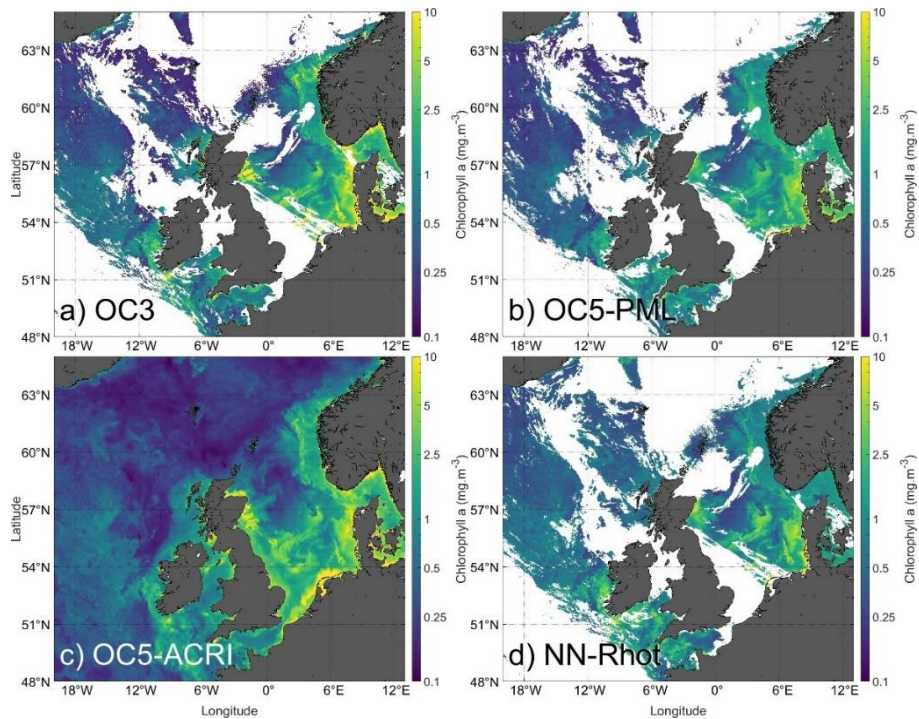

Figure 10: Daily Chlorophyll a surface concentration for the 20th of April, 2005 from a) OC3, b) OC5-PML, c) OC5-ACRI and d) the median of 10 neural networks using Rhot applied to MODIS Aqua. For a) and d), MODIS Aqua images at 12:15, 13:50 and 13:55 were merged. For c), the image was interpolated using +/- 30 days by ACRI-ST to get a cloud-free product.

Figure 11 shows the daily Chl image from 31st December 2019 (same as Figure 5). Given the latitude and time of year, this image represents an example of algorithm performance for high solar zenith angle. Both OC5-PML and OC5-ACRI products apply additional solar zenith angle flags which limit data availability at higher latitudes in winter, with the PML product in this case offering no data availability, while ACRI is using a visible solar zenith angle threshold. In contrast, both OC3 and the NN-Rhot provide data across the scene having been produced from a more permissive dataset using only the ATM FAIL mask and not applying any solar or sensor zenith angle threshold. The OC3 and OC5-ACRI algorithms return significantly higher values than NN in many areas in the southern North Sea and various other coastal waters that are known to present higher sediment loads at this time of year due to winter mixing in shallow waters, or which are generally tidally mixed. The NN produces lower values which are more consistent with previously measured distributions in this region e.g. usually below 0.5 mg.m-3 at Stonehaven [70] (57°N, 2°W) or in the English Channel [71]. Overall, it seems likely that the NN product is both outperforming and is more available in winter than the other direct observation algorithms tested here (OC5-ACRI provides more data points through the wide time frame used, but is therefore not an entirely direct observational algorithm). Whilst the general performance of the NN-Rhot algorithm is reasonably well documented above (e.g. Figure 8d and 9d), we note the occurrence of relatively high Chl values (between 1 and 5 mg.m-3) in a number of coastal areas including the Solway Firth, Morecambe Bay and the Wash. These are regions of known high turbidity and also extensive mudflats at low tide. Algorithm performance under these conditions remains uncertain. Moreover, winter and low values are underrepresented in this dataset, the main limit for machine learning algorithm development. Indeed, further direct validation effort is required for the most optically complex waters and other challenging situations such as areas affected by cloud shadows or immediately adjacent to clouds and land. Masking these areas may be the best option at the moment.

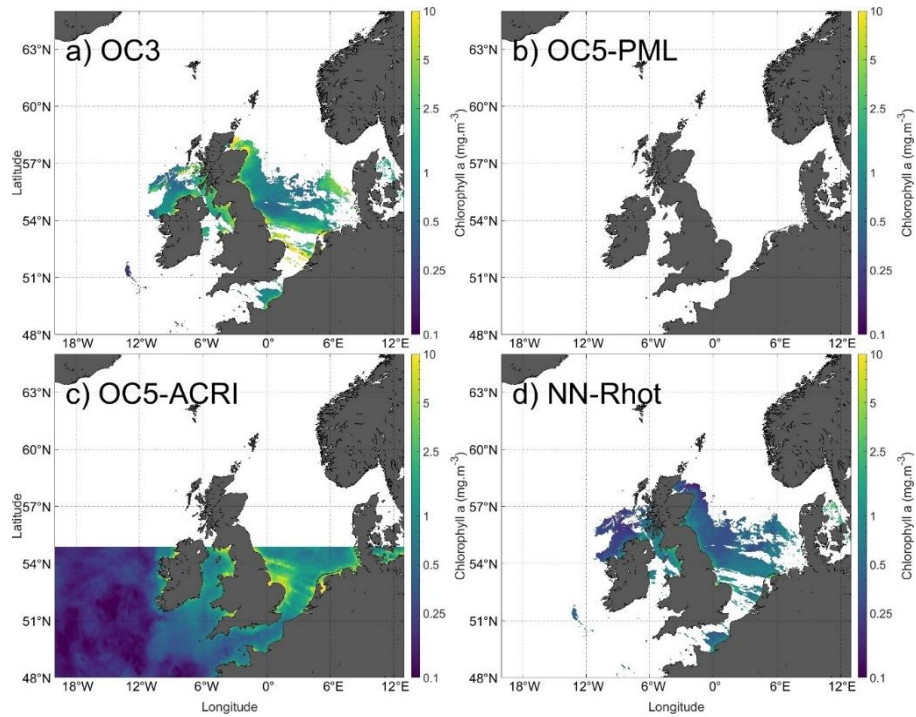

Figure 11: Chlorophyll a surface concentration for the 31st of December, 2019 from a) OC3, b) OC5-PML, c) OC5-ACRI and d) the median of 10 random neural networks using 3 layers of 15 neurons for the Rhot reflectance. Single MODIS-Aqua image at 13:05 for a) and d). Averaged daily image for b). Daily image interpolated (+/- 30 days) for c).

One possibility to independently evaluate performance of the NN algorithm for estimating Chl in coastal waters is the use of a coastal time series. Weekly Chl samples have been collected by Marine Scotland Science since 1997 from the top 10 meters of the ocean at the Stonehaven station (east coast of Scotland). These samples have been co-located with satellite products. Stonehaven matchups were not used for the training, hence their estimation is totally independent. Performances for this specific coastal time series (Figure 12) are slightly worse than the global dataset for both PML and NN products, but statistical metrics on such low amounts of data may not be fully representative. The OC5-ACRI product is not shown as it did not show seasonal correlation at any time and produced significant over-estimates most of the time. 81 matchups are available for the PML product, and slightly more than 2.5 times more for the permissive MODIS Aqua product with 214 daily averaged samples. Data availability comparison between both products is similar to previously observed values for the full dataset (2.8 times more data for MODIS Aqua permissive approach). Compared to the NN, winter data are underrepresented in the PML product due to application of solar zenith angle flags, with solar zenith angles commonly above 70° during winter at this latitude (57°N). Matchups for OC5-PML are available from the middle of February to the middle of November; while they are available at any time using the NN-Rhot algorithm. Low Chl values (<1 mg.m-3), usually sampled between October and March, tend to be overestimated by both algorithms, but the discrepancy tends to be much lower for the NN-Rhot algorithm. NN-Rhot produces consistent estimations with independent *in situ* observations (Figure 12d). General performance metrics for this independent data set are broadly comparable with the original training data set (Figures 8d and 9d). Low values are still systematically overestimated, as a result of under representation of such data from the training dataset. This observation is supported by Figure 12e and 12f that respectively display the standard and relative standard deviation for NN-Rhot estimates of Chl using the ensemble approach. The relative standard deviation is typically greatest in winter and with values reaching ~50%, whereas values drop to ~20% at other times of the year. Again, this reflects scarcity of training data from winter months due to increased cloud cover, reduced daylight hours and potentially reduced sampling effort at this time of year.

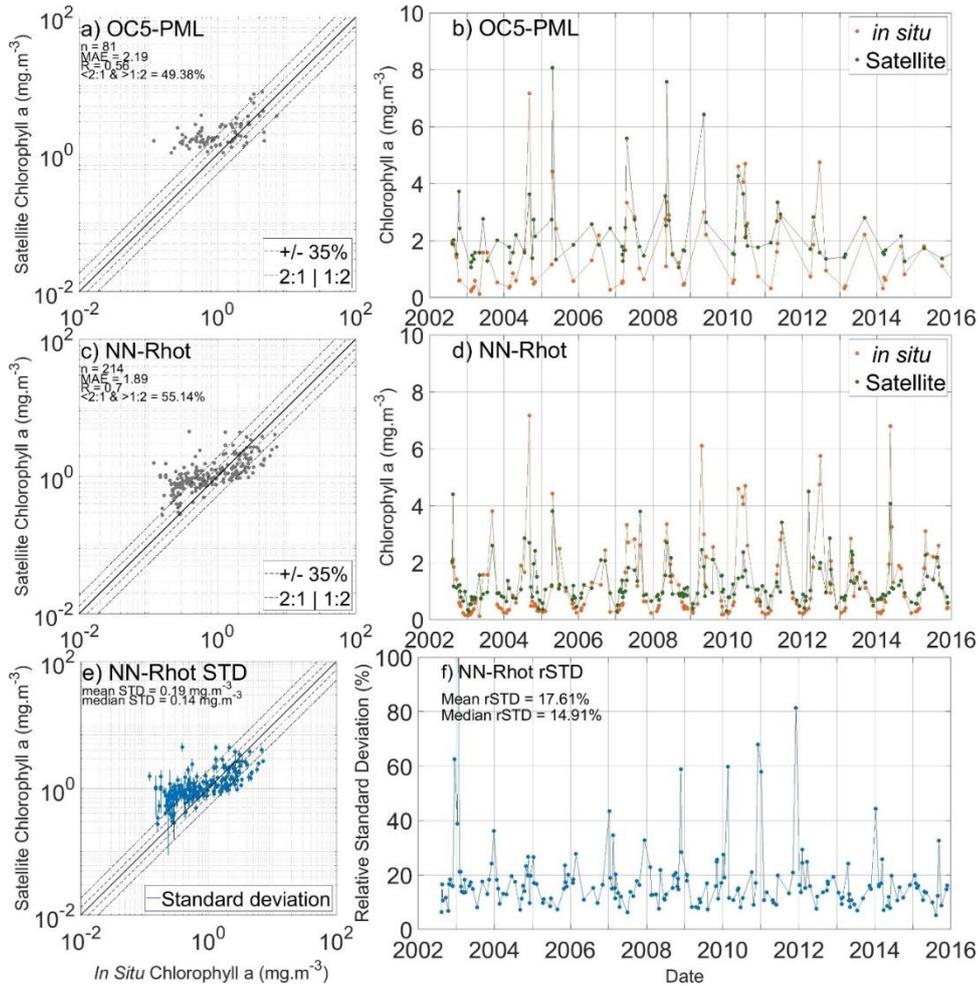

Figure 12: Chlorophyll estimates for the Stonehaven time series, for a) and b) OC5-PML algorithm; and c) and d) NN-Rhot. Standard deviation and relative standard deviation from the ensemble NN-Rhot approach (e and f).

## 4. Discussion

It has been shown that the performance of several blue-green reflectance ratio algorithms, including the latest versions of the most up to date variants, have variable performance in the optically complex waters of the northwest European shelf. At the same time, there is growing interest in the potential use of OCRS to contribute to monitoring the environmental status of territorial waters, particularly with respect to the impacts of eutrophication. There is therefore a profound need to understand and quantify the performance characteristics of satellite Chl products and, if possible, to improve upon the existing array of algorithms. There is growing appreciation of the potential for machine learning approaches to be the key to improved data quality. Here, we have shown that an artificial neural network is able to return favourable performance compared to existing algorithms in direct matchup analyses and can produce realistic images of Chl distributions across the region. It seems likely that efforts of this nature will continue to be developed in coming years and that there will indeed be significant advances in Chl algorithm performance as a result.

In developing a new algorithm like this, it is important to appreciate where previous algorithms have struggled and to understand the limitations of new techniques. It is also important to understand the quality of data being used for validation. Here we are using a set of *in situ* Chl observations including filtered water samples (analysed using HPLC and fluorometric analyses) and *in situ* fluorometry. Of these, HPLC is probably the most accurate source of data. [54] demonstrated that with suitably stringent quality control measures in place, it was possible to achieve average uncertainties below 10%. However, this represented a round-robin exercise by leading laboratories rather than general community capability, and the metric that really matters in this analysis is the range of uncertainty, which was ~±20% for the [54] study. More recently [57] and [72] have both reported Chl results that suggest greater levels of uncertainty for HPLC data sourced from across the community, with uncertainty ranges up to ~±40%. When considering the performance of satellite Chl products, particularly with respect to the potential for using satellite data to complement *in situ* observations for reporting against legislative requirements, it is very important to consider that the quality of the *in situ* data is likely to be of the order of ±40% at best, and that in many cases it may be considerably more uncertain than that. Results from the NN-Rhot algorithm may well be reaching levels that are consistent with the quality of data being used to train the networks, at least for scenarios that have been sampled relatively frequently. That said, there is undoubtedly a need to target future sampling efforts towards scenarios that are currently under-sampled e.g. low Chl in winter, open Case 1 waters and specific events such as coccolithophore blooms.

4.1. Chl algorithm and atmospheric correction failure in northwest European shelf seas.

Northwest European shelf seas are optically diverse, with optical properties ranging from Case 1 conditions at the margins and in certain summer stratified shelf waters to highly tidal regions characterised by high sediment loads and areas of freshwater influence where high concentrations of CDOM impact strongly on reflectance signals [52]. Many areas show strong seasonal variations, with shallow regions often exhibiting higher sediment loads in winter associated with increased wind-driven mixing associated with winter storms [73]. The occurrence of sediment dominated waters degrades the performance of standard atmospheric correction algorithms based on the black pixel approximation. At its extreme, this leads to negative reflectance values in AC-corrected Rrs values, as seen in Figure 5b. However, it is important to realise that over correction is not restricted to spectra with negative values and that many of the non-negative spectra shown in that figure will also be poor representatives of the true remote sensing reflectance signal at sea level. A further measure of the true difficulty of atmospheric correction over these waters is revealed in Figure 5d, where application of just the first stage of atmospheric correction, the Rayleigh correction, is sufficient to drive a number of Rhos spectra into negative values. It is clear that, for an optically complex area like northwest European shelf seas, atmospheric correction is a potentially limiting step. Variable atmospheric correction performance will almost certainly have deleterious consequences for subsequent Chl algorithm performance. Several atmospheric correction methods exist ([3,6-8,59,65). However there is no generally agreed, optimal choice of atmospheric correction that can be successfully applied over such a complex area. More recently, similar observations of higher performances has been made for either temperature estimations from the visible spectra [47] or land observations [74], promoting the use of uncorrected top of atmosphere signal until more performant atmospheric correction methods emerge. The fact that the NN-Rhot algorithm operates successfully from TOA reflectances without requiring selection of an atmospheric correction algorithm is therefore an attractive feature of the approach.

Moving beyond atmospheric correction issues, it has previously been established, using *in situ* radiometry (therefore unaffected by AC issues) that the presence of independent and variable concentrations of sediment and / or CDOM has potential to disrupt the performance of OCx blue-green reflectance ratio algorithms. As one of many examples in the literature, [16] demonstrated that the OC4v4 algorithm performed poorly across the Irish Sea (part of this study area) with notably worse performance in highly turbid, sediment-rich waters. Thus the second stage of Chl estimation from radiometry, the Chl algorithm itself, is further challenged by the level of optical complexity found in northwest European shelf seas. The extremely variable performance of OC3 (Figure 8a) is probably most associated with the performance of the empirical Chl algorithm rather than the AC (similar overestimations with the more restricted version, Figure 9a). The more sophisticated OC5 algorithm version from PML reduces error to within an order of magnitude, although some of this is achieved by masking out identifiably poor quality reflectance data rather than producing inherently better estimates. [75] have shown limitations of this specific OC5-PML algorithm when used for high Chl coastal waters. In both cases, the failure of state of the art algorithms is not only due to atmospheric correction, but is intrinsic to the performance of the algorithms for optically complex waters.

There is considerable interest and optimism in the field that machine learning techniques can be used to develop a new generation of ocean colour algorithms that will perform more robustly in optically complex shelf seas. In this study, we have attempted to develop a baseline approach where we take into account the issues affecting both atmospheric correction and Chl algorithm performance, and where we seek to establish performance characteristics for one of the simplest forms of machine learning techniques. Here we have attempted to systematically explore the various decisions that go into constructing a NN. Notably this has included the option of using any of fully corrected BOA Rrs, partially corrected Rhos and completely uncorrected TOA Rhot as inputs. Remarkably, we have shown that similar levels of performance can be obtained with any of these input data sets, with uncorrected TOA Rhot data providing marginally superior results than the other two. This otherwise surprising result can be explained by the limited performance of AC for these conditions. It is true for both NNs and other, more traditional algorithms, that poorly AC-corrected Rrs data is a hurdle to be overcome. In this case it would appear that eliminating AC altogether and operating directly on Rhot facilitates the job of deriving Chl for the NN.

Advanced algorithms such as OC5 undoubtedly do a more robust job of directly dealing with optical complexity found in coastal waters than the OCx algorithms designed for Case 1 waters. However it is clear from results presented here that a significant portion of the apparent improvement in performance is derived from the relatively stringent flagging used to eliminate the most problematic scenarios. The OC5 –PML product is a good example of a trade-off between improved data quality vs reduced data availability. Conversely, the NN-Rhot approach has been designed to maximise both data quality and availability simultaneously. The computational flexibility offered by the NN allows us to operate directly on TOA reflectances directly and to accommodate the optical complexity of north-western European shelf seas. In doing so the NN-Rhot approach is able to improve both data quality and quantity, and through the ensemble approach it can also provide a measure of data uncertainty. At this point in time the availability and comprehensiveness of the training data set appears to be the limiting factor for the NN-Rhot approach. Further extension of the training data set is perfectly feasible through data mining existing historical data and targeted future sampling.

Another important feature of the NN approach presented here is the use of all relevant spectral bands. Rather than attempting to find an optimum set of wavebands or trying to ascribe physical significance to any particular band, our approach has been able to provide the NN with all available wavebands in the Vis-NIR-SWIR range in order to allow it to resolve the combined problem of dealing with AC and Chl retrieval. The NN approach developed here effectively ingests all of the available spectral information and the neural network is free to determine statistically robust relationships free from human intervention or bias. For example, reducing the number of inputs to the 3 RGB bands (490, 550, 670 nm) used by [34] produced significantly poorer quality results (not shown). It is likely that our approach does in fact carry elements of redundancy e.g. using the 859 and 868 nm or the 547 and 555 nm bands simultaneously may not be meaningful as the input information content is presumably almost identical within each pair. This may point to further simplification that could improve computational efficiency in the future, but is unlikely to improve product quality. We note that [76] and [77] identify failure of the atmospheric correction and resulting impact on blue BOA Rrs wavebands as major limiting factors in their incorporation in NN to retrieve Chl from satellite data. This is consistent with our observations in Figure 5 and our interpretation of why the NN operating on TOA Rhot outperforms the NN operating on BOA Rrs. The inclusion of Red - NIR – SWIR bands, elsewhere used for AC, has an unresolved, but potentially crucial role for successful exploitation of TOA Rhot as input as they can be directly linked to in water sediment concentrations. Recent work supports the need for NIR bands for high Chl content that blue-green reflectance ratio algorithms have problems with [75]. The neural network approach developed in this study uses a combination of visible and infrared bands and there is potential for sensitivity to fluctuations in atmospheric signals such as impacts of volcanic eruptions [78]. We have not observed such impacts in this study, but caution that there remains scope for this to occur under specific circumstances.

4.2. Incorporation of non-optical information to improve NN performance.

The ability of NNs and other machine learning approaches to derive statistically meaningful relationships for seemingly poorly or uncorrelated data is one of the major attractions of the approach. However, there are potential pitfalls that one must also be aware of. An obvious source of potentially useful additional information would be inclusion of geo-spatial and temporal information in the training data set to facilitate recognition of regional and seasonal / inter-annual variations. Inclusion of latitude, longitude, day of year or season as inputs in addition to reflectance signals was attempted and found to produce significantly improved matchups (+10%). However, whilst the associated statistical metrics were improved, it rapidly became clear that the resulting NNs were much weaker in terms of generalisation, with resulting images showing much less spatial detail than would be expected for this region, for example predicting smooth features over large areas in the Atlantic hiding the mesoscale features associated with surface Chl.

It seems likely that inclusion of geospatial and temporal data as inputs to the training data set allowed the NN to identify key features of the data set, but reduced the weighting put onto the directly observed light data. Inclusion of geospatial data, in particular, is likely to overstate regional attributes and to seriously impinge on performance away from areas of dense *in situ* sampling. The NN presented in this paper is a regional algorithm as a consequence of the nature of the training data set used to develop it. However, as an all-optical algorithm, the methodology we have presented has scope to be scaled up to global scale simply by accumulating a sufficiently extensive and robust training data set. When this is attempted, it will be interesting to test the effect of including geo-spatial information as an additional input (potentially attractive if the training data is truly global). Based on our experience to date, we expect that such an approach would also result in a loss of spatial resolution in resulting images, and that future NNs should focus on use of all-optical inputs to maximise generalisation capability. Inclusion of solar and sensor zenith angle or log transformation of inputs prior to normalisation were also tested but only returned slightly improved performances (<2%) and were not used. They could potentially be more impactful at a global scale.

4.3. Benefits and limits of Neural Networks.

NN approaches are generally assumed to be computationally intensive. However, modern computing power is such that all of the work presented here was easily achieved on a relatively modest computing platform. Testing of all the architectures presented in Figure 3 took approximately 1 day using a desktop PC with 16Gb RAM and Matlab R2020a. On the same computer, applying the 10 iteration NN to the image shown in Figure 7a took ~3 minutes. Whilst the computational requirements should not be underestimated, it seems quite feasible that the NN approach developed here is practically implementable due to their low dimension.

In this study we have used the NN as a black box, deliberately avoiding introducing user bias into the production of the NN, though noting that there are inevitable elements of user choice in the design of the NN e.g. normalisation method [79], choice of activation function, choice of error metrics used to assess performance that do in fact impact on eventual NN performance. Unfortunately, the resulting statistical relationships that emerge from NN development are not amenable to physical interpretation. The 3 layers of 15 neurons architecture adopted in this study, despite being relatively small, still represents approximately 690 weights connecting the neurons and is therefore essentially impossible to interpret physically. It should be noted that there is tremendous potential for further refinement of the NN structures, for example with inclusion of dropout layers [80], use of more complex activation functions such as leaky ReLU [81] or different neural network architectures (e.g. long-short term memory networks, [82]). The simple feed forward networks used here provided good performances and whilst further complexity is possible, there is perhaps merit in minimising the complexity of NN structures used and addition of further complexity should be based on demonstrated merit only.

Whilst the NN approach presented here operates on TOA reflectance data with minimal flagging, masks are still applied for clouds, ice, glint, saturation and stray light. It is interesting to note that neural networks have already demonstrated good potential to identify these areas [83] and could replace current threshold methods in the near future. The TOA NN does not require additional ancillary data products and is therefore independent of availability of other data sources. However, there is an opportunity to include these ancillary data that impact the light signal of the atmosphere, and could lead to further improvement.

A novel feature of our approach is the development of a bootstrap-like, iterative approach to produce distributions of Chl estimates for each pixel rather than a single value. The resulting descriptive statistics are potentially useful for providing end users with estimates of confidence in each pixel and for identifying water quality scenarios that are under-represented in training data sets. This can be used to direct future *in situ* sampling efforts to maximise impact on development of future versions of NN algorithms.

NN performance is ultimately determined by how representative the training data set is. For example, in this study, there are only a few hundred samples available for the NE Atlantic and NN performance is currently questionable for that region and for open ocean waters more generally. The focus of this work was to develop an algorithm that worked well in optically complex shelf seas rather than open waters where standard algorithms such as OC3 and CI are expected to work reasonably close to the mission target of +/-35% [84], with MADs of 1.4 for oligotrophic to 1.6 for general case 1 waters usually reached by these algorithms [64]. More generally, the training data set assembled for this study has relatively small numbers of data outside the 1-10 mg.m-3 range (Figure 1b) with potential implications for NN performance towards both extremes of the data range (Figure 8d). Further expansion of the training data set is imperative, particularly if the NN will be used for open ocean waters, even inadvertently as we have done in this paper. Recent work by [20] highlights the potential for machine learning approaches to simplify Chl retrieval in open ocean Case 1 and oligotrophic waters. However, it is less clear if the NN will be able to recover very low values in more turbid waters, a scenario that our current data set does not properly encompass. It is possible that the contribution to reflectance of a low concentration of phytoplankton amongst a high concentration of sediment is sufficiently small that it is not identifiable, even using a NN (e.g. [28]). Moreover, adjacency effects from land is significant in coastal waters [85]. Development of machine learning enhanced OCRS algorithms is likely to increase rather than decrease demand for *in situ* validation data, with particular emphasis on directing effort towards novel and rare features that are underrepresented in existing training data sets.

As it currently stands, the NN developed here is unashamedly a regional algorithm, with our focus being on establishing a methodology that can in future be scaled up to global levels. Testing NNs performances on independent datasets is a major limit as matchup datasets created in the past for other studies typically did not use Rhot nor the full set of bands used for this study. Here we have tested the NN-Rhot approach on an independent coastal time series, with the NN returning encouraging results across the annual cycle. This study highlights the need for agreed matchup datasets to be shared and used by the community for algorithm development, including top, middle and bottom of atmosphere reflectances and the full set of flags. OC5-PML and OC5-ACRI products would benefit from having access to the data gathered for this study to further refine their LUTs, and could potentially return higher performances. However, there would still be limits associated with the nature of the OC5 algorithm and reliance on flagging to eliminate the more difficult scenarios.

Restricting NN inputs to optical signals only is potentially key to ensuring translation of the NN approach beyond current geographic confines. However, there is also a limitation on applicability to a particular satellite sensor, in this case MODIS Aqua. This is partly due to the availability of specific bands for each sensor but also reflects specifics of sensor calibration. Directly translating the current NN to another sensor is unlikely to be easy and is likely to require collation of a suitable matchup data set for that instrument followed by repetition of the methodology outlined above generating another instrument-specific NN. Development of a long term, consistent TOA time series, incorporating data from multiple satellites along the lines of the OC-CCI project, is essential and in this case may be key to developing a global data set for exploiting the capabilities offered by machine learning data analysis techniques in this field. Introduction of hyperspectral OCRS data in the future, e.g. the forthcoming Plankton, Aerosol, Cloud, ocean Ecosystem (PACE mission, https://pace.gsfc.nasa.gov), has potential to support development of improved NNs that may be able to exploit enhanced spectral resolution to improve accuracy of Chl retrieval.

The NN developed here proceeds straight from TOA Rhot to estimates of Chl, effectively bypassing the need for production of atmospherically corrected, BOA Rrs values. Whilst this is efficient, it precludes the possibility of applying the NN to reflectance signals measured *in situ*. Ironically, the Rrs NN discussed in Figure 4 might not work well with *in situ* Rrs data as a result of having been trained on poorly AC corrected satellite Rrs values. Whilst there is clearly merit in avoiding the need for atmospheric correction, there is undoubtedly interest in generating accurate BOA Rrs values, not least because it is a Global Climate Observing System established Essential Climate Variable, but also because it facilitates functional links between satellite and ground truth optical observations. The NN methodology development proposed here is translatable to deriving surface Rrs values instead of Chl, but requires provision of an adequate training data set. Extensive efforts to produce global sets of *in situ* optical and biogeochemical data have been made by the community (e.g. NOMAD [86]; MERMAID [87]) and it is likely that future AC algorithms will be developed using neural networks techniques such as the NN approach discussed here [36,88]. Again, there will be increasing value to be had from future *in situ* optical sampling, with increasing focus on the use of sensors deployed on moorings and other autonomous platforms providing an efficient means of generating necessary matchups with satellite data.

## 5. Conclusion

A methodology has been developed to find optimal artificial neural network architectures for the estimation of Chl in coastal waters from the MODIS Aqua ocean colour sensor using all available visible and short infrared bands related to ocean or atmospheric features. The use of top-of-atmosphere uncorrected reflectance, Rhot, is shown to be feasible using neural networks. For northwest European shelf seas, the neural network algorithms clearly outperformed state-of-the-art ocean colour algorithms for a matchup data set covering the whole MODIS-Aqua era, from July 2002 to January 2020. They returned significantly higher Pearson Correlation (R >0.7 compared to 0.61) and lower Mean Absolute Difference, <1.8 against 2.10) without application of additional data quality flags, thus simultaneously increasing the number of available matchups and the number of pixels per image. As a result, the networks presented here are capable of producing promising quality data in winter when other algorithms are masked out. By operating on Rhot, the network eliminates the need for atmospheric correction which is shown to perform poorly in many instances for this region. Chlorophyll maps are therefore produced with minimal data processing steps, although the application of only a small number of masks to remove non-water or atmospherically impacted areas is still required. Iterative re-sampling of the training data set was used to produce an ensemble of NNs that in turn provide both median best-estimates and uncertainty distributions for each pixel. The addition of geo-spatial and temporal information is discussed but was found to harm neural network performances by transforming them into statistical modelling tools rather than observation tools. The current version of the algorithm is restricted in geographical scope by the extent of the available training data set, but the methodology presented has potential to be upscaled to a global algorithm upon generation of a suitably extensive training data set. There is further potential to adapt the methodology to produce a future neural network that can be applied to merged ocean colour data sets and to use the technique to develop other useful products, including a more robust atmospheric correction algorithm. In all cases, the advent of machine learning based ocean colour algorithms means there is a strong imperative to continue, and if possible expand, *in situ* observation programs that will provide the training data sets needed to update and further improve this type of algorithm.

**Author Contributions:** Madjid Hadjal carried out the data gathering, preparation and practical implementation of the neural network approach. Encarni Medina-Lopez and Jinchang Ren provided expert guidance on development of neural network architectures. Alejandro Gallego provided expert guidance on end user requirements for operational use of ocean colour products. David McKee provided expert guidance on ocean colour remote sensing and supervised development of the methodology for this application.

**Funding:** Madjid Hadjal was supported by a joint MASTS (Marine Alliance for Science and Technology Scotland / Datalab studentship. David McKee was supported by UK Natural Environment Research Council grants NE/E013678/1 and NE/S003517/1.

**Data Availability Statement:** All data underpinning this publication are openly available from the University of Strathclyde KnowledgeBase at https://doi.org/10.15129/32604fa3-ed09-4b5a-b735-9321ddaa8ef9.

**Acknowledgments:** This study has been conducted using E.U. Copernicus Marine Service Information. The authors would like to thank Francis Gohin, Ifremer (French Research Institute for Exploitation of the Sea, France) for providing details about the OC5 algorithm and James Harding (Edinburgh University) for help setting up the networks and useful discussions. The authors thank the International Council for the Exploration of the Sea (ICES) and Marine Scotland Science (MSS) for facilitating this research by providing merged data sets. Thanks to A. Holbach from Department of Ecoscience, Aarhus University (Denmark) for accessing and providing data from the Danish National Monitoring Programme; Dr Kjell Gundersen and the Plankton Research Laboratory at IMR, for access to surface Chl-a data from the Norwegian and Barents Seas; Tim Smyth for providing PML time series at E1/L4 stations, and Naomi Greenwood and Michelle Devlin for providing access to Cefas SmartBuoy data that were collected under the UK Department for Environment, Food and Rural Affairs (Defra) contracts SLA25 and GiA03. The authors thank NASA for making available the MODIS Aqua data and the Seadas data processing software. The colour-blind friendly colormaps used for this study were generated using the MatPlotLib 2.0 library and we thank their contributors. All the figures and data analysis used for this study have been done using Matlab R2020b and we thank their contributors.

**Conflicts of Interest:** The authors declare no conflict of interest.